\documentclass[onecolumn,12pt]{article}
\usepackage{graphicx}
\newcommand\simgt{\hspace{0.3em}\raisebox{0.4ex}{$>$}\hspace{-0.75em}\raisebox{-.7ex}{$\sim$}\hspace{0.3em}}

\begin{document}
\baselineskip 14pt

\title{Distortion of Magnetic Fields in a Starless Core V: \\
Near-infrared and Submillimeter Polarization in FeSt 1-457} 
\date{Ver.2}
\author{Ryo Kandori$^{1}$, Tetsuya Nagata$^{2}$, Ryo Tazaki$^{3}$, Motohide Tamura$^{1,4,5}$, \\
Masao Saito$^{4}$, Kohji Tomisaka$^{4}$, Tomoaki Matsumoto$^6$ Nobuhiko Kusakabe$^{1}$, \\
Yasushi Nakajima$^{7}$, Jungmi Kwon$^{8}$, Takahiro Nagayama$^{9}$, and Ken'ichi Tatematsu$^{4}$\\
{\small 1. Astrobiology Center of NINS, 2-21-1, Osawa, Mitaka, Tokyo 181-8588, Japan}\\
{\small 2. Kyoto University, Kitashirakawa-Oiwake-cho, Sakyo-ku, Kyoto 606-8502, Japan}\\
{\small 3. Astronomical Institute, Graduate School of Science Tohoku University,}\\
{\small 6-3 Aramaki, Aoba-ku, Sendai 980-8578, Japan}\\
{\small 4. National Astronomical Observatory of Japan, 2-21-1 Osawa, Mitaka, Tokyo 181-8588, Japan}\\
{\small 5. Department of Astronomy, The University of Tokyo, 7-3-1, Hongo, Bunkyo-ku, Tokyo, 113-0033, Japan}\\
{\small 6. Faculty of Sustainability Studies, Hosei University, Fujimi, Chiyoda-ku, Tokyo 102-8160}\\
{\small 7. Hitotsubashi University, 2-1 Naka, Kunitachi, Tokyo 186-8601, Japan}\\
{\small 8. Institute of Space and Astronautical Science, Japan Aerospace Exploration Agency,}\\
{\small 3-1-1 Yoshinodai, Chuo-ku, Sagamihara, Kanagawa 252-5210, Japan}\\
{\small 9. Kagoshima University, 1-21-35 Korimoto, Kagoshima 890-0065, Japan}\\
{\small e-mail: r.kandori@nao.ac.jp}}
\maketitle

\abstract{
The relationship between submillimeter (submm) dust emission polarization and near-infrared (NIR) $H$-band polarization produced by dust dichroic extinction was studied for the cold starless dense core FeSt 1-457. The distribution of polarization angles ($90^{\circ}$-rotated for submm) and degrees were found to be very different between at submm and NIR wavelengths. The mean polarization angles for FeSt 1-457 at submm and NIR wavelengths are $132.1^{\circ} \pm 22.0^{\circ}$ and $2.7^{\circ} \pm 16.2^{\circ}$, respectively. The correlation between $P_H$ and $A_V$ was found to be linear from outermost regions to relatively dense line of sight of $A_V \approx 25$ mag, indicating that NIR polarization reflects overall polarization (magnetic field) structure of the core at least in this density range. The flat $P_H/A_V$ versus $A_V$ correlations were confirmed, and the polarization efficiency was found to be comparable to the observational upper limit (Jones 1989). On the other hand, as reported by Alves et al., submm polarization degrees show clear linearly decreasing trend against $A_V$ from $A_V \approx 20$ mag to the densest center ($A_V \approx 41$ mag), appearing as \lq \lq polarization hole'' structure. The power law index for the $P_{\rm submm}$ versus $A_V$ relationship was obtained to be $\approx -1$, indicating that the alignment for the submm sensitive dust is lost. These very different polarization distributions at submm and NIR wavelengths suggest that (1) there is different radiation environment at these wavelengths or (2) submm-sensitive dust is localized or the combination of them. 
}

\vspace*{0.3 cm}

\clearpage

\section{Introduction}
Magnetic filed is thought to play an important role in the formation and evolution of molecular clouds, its cores, stars, and planetary systems (e.g., Crutcher 2012; McKee \& Ostriker 2007). Measurements of polarizations produced by dust grains are one of the most powerful tools to probe into magnetic field geometry and strength at various wavelengths. According to the general alignment mechanism of interstellar dust grains, the elongated dust grains spin about their axis (minor axis) oriented parallel to the magnetic fields (Andersson et al. 2015 for review). Thus, dichroic extinction by dust often observed at optical to near-infrared (NIR) wavelengths can produce incident polarization vectors parallel to the plane of sky magnetic fields, and dust emission polarimetry at submillimeter (submm) to far-infrared (FIR) wavelengths can obtain polarization vectors perpendicular to the plane of sky magnetic fields. 
\par
Though magnetic field structure can be traced at various wavelengths, different observing wavelengths are sensitive to the different size of dust: e.g., optical polarimetry is sensitive to small size grains and submm polarimetry is sensitive to large size grains. Moreover, the probing depth in column density is different at each wavelength. To understand magnetic field structure as well as dust grain properties, polarimetry at multiple wavelengths is essential. 
\par
Dust emission polarimetry, particularly at submm to FIR wavelengths, has proven to be a powerful technique for tracing magnetic field structures in regions of high column density, such as in giant molecular clouds or dark cloud complexes (e.g., Houde et al., 2004; Matthews et al. 2002; Sugitani et al. 2010) and protostellar envelopes (e.g., Girart et al. 2006, Rao et al. 2009). Unlike the starlight polarimetry, the dust emission polarimetry can probe the (dense) region exactly where dust is emitting. However, it is also known that the \lq \lq polarization hole'' effect, i.e., the phenomenon that the weakest polarization degree is observed toward the densest central regions, exists in almost every emission polarimetry image (e.g., Matthews et al. 2009; Hull et al. 2014; Brauer et al. 2016). Jones et al. (2015) found that for starless cores in the range of $A_V \simgt 20$ mag, the slope of the $P_{\rm submm}$ versus $\tau$ becomes $\sim -1$, indicating that the grain alignment is lost beyond the depth. Thus, it is not valid that the submm or FIR polarizations reflect the magnetic field structure of overall clouds or cores. 
\par
There are studies of the comparison between submm or FIR emission polarimetry with optical or NIR dichroic extinction polarimetry. Kandori et al. (2007) conducted NIR polarimetry ($H$ band) of the NGC 2024 region. Though the comparison between NIR polarization vectors and those at 100 $\mu$m and 850 $\mu$m with $90^{\circ}$-rotation results in good correlation for the relatively diffuse region ($A_V < 50$ mag), the correlation becomes bad toward the highly obscured molecular ridge region ($A_V \ge 50$ mag). This may be due to the limitation of probing depth at NIR wavelegths, typically $A_V < 50$ mag, or the effect of the polarization hole. Kusakabe et al. (2008) conducted similar studies for OMC-1 (M42) region, and obtained results that the direction of magnetic fields obtained at NIR ($H$ band) wavelengths is identical to those at 350 $\mu$m. Since M42 is not the region deeply embedded in the parent molecular cloud, the effect of limited probing depth at NIR wavelengths does not seem severe. Santos et al. (2017) conducted $I$ band polarimetry toward Vela-C molecular cloud and compared it with the 500 $\mu$m dust emission polarimetry data. The obtained magnetic field directions are generally consistent with each other. 
\par 
The most detailed comparison study has been conducted by Alves et al. (2014, 2015) for the starless dense core FeSt 1-457. FeSt 1-457 is cataloged as a member of the the Pipe Nebula dark cloud complex (as core \#109 in Alves, Lombardi, \& Lada 2007; Onishi et al. 1999; Muench et al. 2007), located in the direction of the Galactic center, at a distance of $130^{+24}_{-58}$ pc (Lombardi et al. 2006). Owing to the core's relatively isolated geometry, simple shape, and plenty background stars, a density structure study using the Bonnor--Ebert model (Bonnor 1956; Ebert 1955) was conducted based on the measurements of dust extinction at NIR wavelengths (Kandori et al. 2005). The physical properties of the core are well determined with a radius of $18500 \pm 1460$ AU (144$''$), mass of $3.55 \pm 0.75$ M$_{\odot}$, and central density of $3.5(\pm 0.99) \times 10^5$ cm$^{-3}$ (Kandori et al. 2005) at a distance of 130 pc (Lombardi et al. 2006). 
\par
In the study by Alves et al. (2014, 2015), optical ($R$ band), NIR ($H$ band), and submm (870 $\mu$m) polarimetry data were compared with each other. This is the first detailed multiple wavelength polarization studies toward a starless core. Though optical and NIR data show consistent polarization vector angles ($168^{\circ} \pm 4^{\circ}$ and $163^{\circ} \pm 5^{\circ}$, measured from north to east direction), unlike the previous studies described above, submm data ($130^{\circ} \pm 12^{\circ}$) systematically bend from NIR data by $\approx 35^{\circ}$ in derived magnetic field direction (Figure 1 of Alves et al. 2014). Moreover, submm polarization degree clearly decreases from outer region to the densest center appearing as a polarization hole. Thus, the submm polarization seems not to reflect the overall polarization structure of the core, and the difference of polarization angle between optical/NIR and submm wavelength may be due to the effect of polarization hole. 
\par
In the analysis by Alves et al. (2014), NIR stellar polarimetry data for the core were the superposition of polarizations from the core itself and ambient medium. They did not isolate the core polarization. Since the background starlight polarimetry measures the integrated dichroic polarization, the subtraction of the polarization from ambient medium is essential. Kandori et al. (2017a), hereafter Paper I, conducted NIR polarimetry of FeSt 1-457 and subtraction analysis of the ambient polarization from the core polarization, and found that the magnetic field structure associated with FeSt 1-457 was hourglass-shaped, which was very different in appearance from the original data (see Section 3.1). The hourglass-shaped polarization distribution was modeled using simple three-dimensional (3D) polarization model (Kandori et al. 2017b, hereafter Paper II). In Kandori et al. (2018a), hereafter Paper III, core polarization was corrected for (1) ambient polarization, (2) depolarization effect of inclined distorted 3D magnetic fields, and (3) magnetic inclination angle toward lines of sight. 
\par
In the present study, these NIR polarization data were compared with the submm polarization data (Alves et al. 2014). The derived magnetic field direction as well as polarization degree distribution of the submm data were found to be significantly different from those at NIR wavelengths. 

% Kandori+2007; Kusakabe+2008, Alves+2014(Juarez+2017), Santos+2017
% Kandori+2017a,b; Kandori+2018a

\section{Data and Methods}
The submm polarimetric data was provided by Alves et al. (2014). Observations were conducted at 345 GHz using the Atacama Path Finder Experiment (APEX) 12m telescope attached with LABOCA bolometer and PolKa polarimeter (for the instrument, see Siringo et al. 2004, 2012; Wiesemeyer et al. 2014). The half power beam width (HPBW) at the wavelength is $\sim 20''$. The 600 scans of 2.5 min each was conducted to obtain in total of 25 hr on source integration time. 
% Each scan consists of four subscans with a spiral stroke pattern with each centered on a different position. 
The instrumental polarization of their data is about $0.10 \pm 0.04$ \% toward the peak of emission. The data was corrected for instrumental polarizations, and the rms noise of the final submm map was typically $\sim 5$ mJy/beam. The 87 data points with $P_{\rm submm} / \delta P_{\rm submm} \ge 2$ were used. 
\par
The NIR polarimetric data of FeSt 1-457 was taken from Paper I. The data was taken using the $JHK$${}_{\rm s}$-simultaneous polarimetric imaging instrument SIRPOL (Kandori et al. 2006, polarimetry-mode of the SIRIUS camera: Nagayama et al. 2003) on the IRSF 1.4-m telescope at the South African Astronomical Observatory (SAAO). SIRPOL is a single-beam polarimeter providing a wide field of view ($7.\hspace{-3pt}'7 \times 7.\hspace{-3pt}'7$ with a scale of 0$.\hspace{-3pt}''$45 ${\rm pixel}^{-1}$). The uncertainty of polarization degree due to sky variation is 0.3\% during each exposure. The uncertainty of the measurements of the origin of the polarization angle of the polarimeter (i.e., determination of the correction angle) is less than $3^{\circ}$ (Kandori et al. 2006 and updates for Kusune et al. 2015). 
% Observations were conducted using the $JHK$${}_{\rm s}$-simultaneous imaging camera SIRIUS (Nagayama et al. 2003) and its polarimetry mode SIRPOL (Kandori et al. 2006) on the IRSF 1.4-m telescope at the South African Astronomical Observatory (SAAO). IRSF/SIRPOL can provide deep (18.6 mag in the $H$ band, $5\sigma $ for one-hour exposure) and wide-field ($7.\hspace{-3pt}'7 \times 7.\hspace{-3pt}'7$ with a scale of 0$.\hspace{-3pt}''$45 ${\rm pixel}^{-1}$) NIR polarimetric data. SIRPOL is a single-beam polarimeter. The uncertainty from the sky variation during exposures is typically 0.3\% in polarization degree. The uncertainty of polarization angle due to the accuracy of the origin of polarization angle of the polarimeter (i.e., determination of the correction angle) is less than $3^{\circ}$ (Kandori et al. 2006 and updates for Kusune et al. 2015).
% The uncertainty of polarization angle is less than $3^{\circ}$ for bright sources (Kandori et al. 2006 and updates for Kusune et al. 2015). 
\par
Figure 1a shows the polarization vector map of point sources toward FeSt 1-457. The background image is the intensity image in the $H$ band. The core located at the center of the image and the white circle denote the core radius ($144''$) determined based on the extinction map study (Kandori et al. 2005). 
% Figure 1a shows the polarization vectors of point sources toward FeSt 1-457, superimposed on the intensity image in the $H$ band. The core appears as a dark obscuration at the center of the image and the radius of the core (144$''$) determined by the density structure study (Kandori et al. 2005) is marked by a white circle. 
\par
Figure 1b shows the polarization vector map of point sources after subtraction of ambient polarization components using the stars located outside the core radius (see, Paper I). The figure is zoomed-in version of Figure 1a showing the region inside the core radius. The number of stars available for tracing the core magnetic fields is 185. In Figure 1b, magnetic fields seems a distorted axisymmetric shape reminiscent of an hourglass. The white lines show the best fitting result of magnetic field lines using a parabolic function $y = g + gC{x^2}$, where $g$ specifies each magnetic field line and $C$ determines the curvature of the magnetic fields. This result serves as the first detection of the hourglass-shaped magnetic fields in dense cores (Paper I). 
\par
% The polarimetry data toward the core is the superposition of polarizations from both the core itself and the ambient medium which is unrelated to the core. The unrelated \lq \lq off-core'' polarization component was spatially fitted in $Q/I$ and $U/I$ in the $H$ band using the stars located outside of the core radius ($R>144''$). The distributions of the $Q/I$ and $U/I$ values are modeled as $f(x,y)=A + Bx + Cy$, where $x$ and $y$ are the pixel coordinates, and $A$, $B$, and $C$ are the parameters to be fitted. 
% \par
% The off-core regression vectors are subtracted from all the polarization vectors in order to isolate the polarization vectors associated with the core. After subtracting unrelated polarization components, 185 stars located within the core radius were selected for further analysis (Figure 1b). 
% \par
% In Figure 1b, the magnetic field follows a distinct axisymmetric shape reminiscent of an hourglass. The white lines show the magnetic field direction inferred from the fitting with a parabolic function $y = g + gC{x^2}$, where $g$ specifies the magnetic field lines and $C$ determines the degree of curvature in the parabolic function. The results represent the first observational evidence of hourglass-shaped magnetic fields in a starless core (Paper I). 
The 3D modeling of the hourglass magnetic fields was conducted (Paper II). In the study, the axially symmetric hourglass field was assumed. The 3D function used to model the magnetic fields was a simple parabolic function $z(r, \varphi, g) = g + gC{r}^{2}$ in cylindrical coordinates $(r, z, \varphi)$, where $\varphi$ is the azimuth angle (measured in the plane perpendicular to $r$). The parabolic function describes the orientation of polarization vectors in 3D, and the degree of polarizations can be obtained from known density structure of the core and the observationally determined polarization-extinction relationship. The polarization vector maps based on the 3D function at various line-of-sight inclination angle are shown in Figure 2 of Paper II. The $\chi^2$ analysis based on the model and the observational data resulted in the line of sight inclination angle $\theta_{\rm inc}$ of $45^{\circ}$ for FeSt 1-457. 
\par
When distorted hourglass-shaped fields are inclined toward the line of sight, the model core can produce depolarization effect, particularly in the equatorial plane of the core (see, Figure 2 of Paper II). The depolarization effect is due to the crossing of the polarization vectors located in frontside and backside of the core. In Figure 2 of Paper II, the depolarization effect is apparent especially in the panels for $\theta_{\rm inc}=30^{\circ}$ and $\theta_{\rm inc}=15^{\circ}$. The depolarization region seems dark patches where polarization degree is low compared with neighboring regions. 
\par
The polarized light from the core's background stars includes several effects as described above. The superposition of ambient polarization components, the depolarization effect due to inclined distorted magnetic fields, and the magnetic inclination angle should be corrected to obtain accurate polarization information of FeSt 1-457. These effect can be corrected using the subtraction analysis and 3D polarization modeling of the core (Paper III).

\section{Results and Discussion}
\subsection{Comparison between NIR and submm polarization}
Though the comparison studies of submm and NIR polarizations have been reported by Alves et al. (2014, 2015), their NIR data were uncorrected with the superposition of ambient polarization, depolarization effect of 3D distorted fields, and line of sight magnetic inclination angle. In this section, we compare the NIR data with appropriate corrections with submm polarization data, and discuss the consistency between these data. 
\par
Figure 2a shows the hourglass-shaped polarization distribution (Paper I) on the $A_V$ distribution measured with $34''$ resolution (Kandori et al. 2005). The peak $A_V$ is $\approx 41$ mag toward the center of the core. The central region with $A_V$ greater than $\approx 25$ mag of the core is too dense to be probed with the current sensitivity of NIR polarization measurements. The plane-of-sky magnetic axis of the hourglass field is $179^{\circ}$, and the polarization distribution is clearly in north-south direction, which is perpendicular to the elongation of the densest region of the core, consistent with (1) the mass accumulation along the magnetic field lines (e.g., Galli \& Shu 1993a,b) or (2) the magnetohydrostatic configulation (Tomisaka, Ikeuchi \& Nakamura 1988). 
\par
Figure 2b shows the submm polarization distribution taken from Alves et al. (2014), which was $90^{\circ}$-rotated to show the direction of magnetic fields. The magnetic field directions obtained from the NIR and submm data differ significantly. The submm vectors flow from north-west to south-east, which does not match the north-south magnetic field distribution obtained at NIR wavelengths. Though the regions traced by both wavelengths are different, the orientations of submm and NIR vectors clearly differ in their overlapping regions. Figures 3a and 3b shows histograms of the NIR polarization vectors and those at submm ($90^{\circ}$-rotated). The mean angles and standard deviations of the distribution at NIR and submm are $2.7^{\circ} \pm 16.2^{\circ}$ and $132.1^{\circ} \pm 22.0^{\circ}$, respectively. There is a deviation angle of $\approx 50^{\circ}$ in between these data. The submm data also deviates by $\approx 30^{\circ}$ from the directions of \lq \lq off-core'' vectors obtained at NIR ($165^{\circ}$, Paper I) and at optical ($165^{\circ} \pm 4^{\circ}$, Franco et al. 2010) wavelengths. 
\par
Figure 4a shows the polarization versus $A_V$ relationship at NIR wavelengths, under the corrections of (1) ambient polarization, (2) depolarization caused by 3D distorted fields, and (3) line-of-sight magnetic inclination angle (see, Paper III). The horizontal axis was not from the $A_V$ mapping data in Kandori et al. (2005), but from the $H-K_s$ colors of stars. $A_V$ was calculated using the relation $A_V = 21.7 \times E_{H-K_s}$ (Nishiyama et al. 2008), and $E_{H-K_s} = (H-K_s) - <H-K_s>_{\rm bkg}$, where $<H-K_s>_{\rm bkg} = 0.40$ mag is the average $H-K_s$ color of stars in the same reference field region used in the previous dust extinction study (see Figure 2 of Kandori et al. 2005). 
% where $<H-K_s>_{\rm bkg} = 0.54$ mag is the average $H-K_s$ color of whole stars in the field of view located outside the core radius ($R > 144''$). 
\par
From Figure 4a, it is clear that the relationship is linear, and the linear least-squares fitting result is $P_H = (0.51 \pm 0.03)A_V - (1.72 \pm 0.46)$. The dotted-dashed line shows the observational upper limit of the relationship described using the equation $P_{K,{\rm max}} = \tanh{\tau_{\rm p}}$, where $\tau_{\rm p} = (1-\eta)\tau_{K}/(1+\eta)$ when the parameter $\eta$ is set to 0.875 (Jones 1989). $\tau_{K}$ denotes the optical depth in the $K$ band. The observed slope in the polarization-extinction relationship for FeSt 1-457 seems comparable to that for the upper limit value. 
The linear relationship is reflected in the flat distribution of $P_H / A_V$ versus $A_V$ diagram (Figure 4b), linearly fitted as $P_H / A_V = -0.0035 A_V + 0.4172$ (dashed line). 
% Considering the scatter in Figure 4a, the data below $A_V = 5$ mag were ignored in the fitting. 
% The obtained slope is close to zero, showing the flat and linear relationship on the $P_H$ versus $A_V$ diagram. 
The fitting of $P_H / A_V$ versus $A_V$ data using the power law $P_H / A_V \propto {A_V}^{\alpha}$ (dotted line) resulted in the $\alpha$ index close to zero, $-0.12 \pm 0.11$. The power law fitting result seems identical to the linear fitting line. The dotted-dashed line in the figure shows observational upper limit in the $P/A$--$A$ relationship (Jones 1989). 
\par
As shown in Figures 4a and 4b, the relationship between $P_H$ and $A_V$ is linear for the $A_V$ below $\approx 25$ mag with high polarization efficiency comparable to that for the observational upper limit (Jones 1989). This indicates that the NIR polarimetry traces overall polarization (magnetic field) structure of FeSt 1-457, although the probing depth ($A_V \le 25$ mag) is not sufficient for the densest regions ($A_V \approx 41$ mag). The simplest interpretation is that, at least $A_V \le 25$ mag, the polarization efficiency per dust grains at NIR wavelength is constant from the outer boundary toward the central regions. Alves et al. (2014) reported that there is a kink at $A_V \approx 9.5$ mag in the $P_H / A_V$ versus $A_V$ relationship. Note that they did not conducted any correction to separate core polarizations from ambient polarizations. The kink disappeared after the corrections of (1) described above (Figure 4 of Paper III), which isolate the core polarizations from the effect of ambient medium. 
\par
Figure 4c shows the submm polarization degree $P_{\rm submm}$ versus $A_V$ diagram. The $A_V$ data was taken from Kandori et al. (2005). Unlike NIR, submm polarization degree data clearly shows decreasing trend against $A_V$, fitted as $P_{\rm submm} = (-0.43 \pm 0.08) A_V + (24.2 \pm 2.6)$. In the fitting, the data below $A_V = 23$ mag were ignored. 
% Considering the relatively low signal-to-noise ratio, $P_{\rm submm} / \delta P_{\rm submm} \ge 2$, the data below $A_V = 23$ mag were ignored in the least-squares fitting. 
The steep rise of $P_{\rm submm}$ at $A_V < 23$ mag may be real, and may indicate the steep increase of polarization efficiency at submm wavelengths toward the outer regions. 
The fitting of $P_{\rm submm}$ versus $A_V$ using the power law $P_{\rm submm} \propto {A_V}^{\alpha}$ (dotted line) resulted in the $\alpha$ index of $-1.11 \pm 0.12$. The obtained index close to $-1$ indicates that the alignment of submm sensitive dust is lost inside the core. This is consistent with the value, $-0.92 \pm 0.17$, obtained by Alves et al. (2014), and is also consistent with the observational report for other starless cores (Jones et al. 2015). The distribution of $P_{\rm submm}$ slightly deviates from the power law fitting (dotted line) toward the low $A_V$ region. This may be indicative of the alignment of dust grains in the outer region of the core. 
% Figure 4d shows the $P_{\rm submm}/A_V$ versus $A_V$ diagram. {\bf The relationship was fitted to be} $P_{\rm submm} / A_V = (-0.022 \pm 0.084) A_V + (1.081 \pm 2.598)$. 
% The linear decrease of polarization efficiency toward the center of the core is clear from the Figure 4c, indicating the existence of polarization hole as suggested by Alves et al. (2014). 
\par
Comparing Figures 4a,b and 4c, the difference is dramatical. The polarization degree relationships against $A_V$ at NIR and submm wavelengths are found to be very different in FeSt 1-457. Though the existence of polarization hole at submm may be the sign of the lack of radiation in the starless core according to the radiative torque theory (e.g., Dolginov \& Mitrofanov 1976; Draine \& Weingartner 1996,1997; Lazarian \& Hoang 2007), the linear $P_H$ versus $A_V$ relationship requires relatively strong radiation fields. 
A possible explanation is the difference of radiation environment of FeSt 1-457 at NIR and submm wavelengths. Another possible explanation is that the localization of submm-sensitive dust in the core. As discussed by Alves et al. (2014), large dust in deep core interior does not contribute to the submm polarization and dust in outer region may relatively well traced by submm polarimetry. This scenario may explain large difference of magnetic field directions obtained at NIR and submm wavelengths. If this is true, the polarization measurements at submm wavelengths may not appropriate to probe into the densest region of starless dense cores. With linear correlations against $A_V$, NIR polarization is proven to be a robust and accurate tool to study magnetic fields toward the outer region to the relatively obscured ($A_V \approx 25$ mag) lines of sight. To investigate magnetic field and polarization efficiency in the inner most regions, deep NIR polarimetries with unprecendeted depth using large 8-m class telescopes are planned.

\subsection*{Acknowledgement}
We are grateful to the staff of SAAO for their kind help during the observations. We wish to thank Tetsuo Nishino, Chie Nagashima, and Noboru Ebizuka for their support in the development of SIRPOL, its calibration, and its stable operation with the IRSF telescope. The IRSF/SIRPOL project was initiated and supported by Nagoya University, National Astronomical Observatory of Japan, and the University of Tokyo in collaboration with the South African Astronomical Observatory under the financial support of Grants-in-Aid for Scientific Research on Priority Area (A) Nos. 10147207 and 10147214, and Grants-in-Aid Nos. 13573001 and 16340061 of the Ministry of Education, Culture, Sports, Science, and Technology of Japan. RK, MT, NK, KT (Kohji Tomisaka), and MS also acknowledge support by additional Grants-in-Aid Nos. 16077101, 16077204, 16340061, 21740147, 26800111, 16K13791, 15K05032, and 16K05303.

\clearpage 

\begin{figure}[t]  % \vspace*{-2.0 cm}
\begin{center}
 \includegraphics[width=6.5 in]{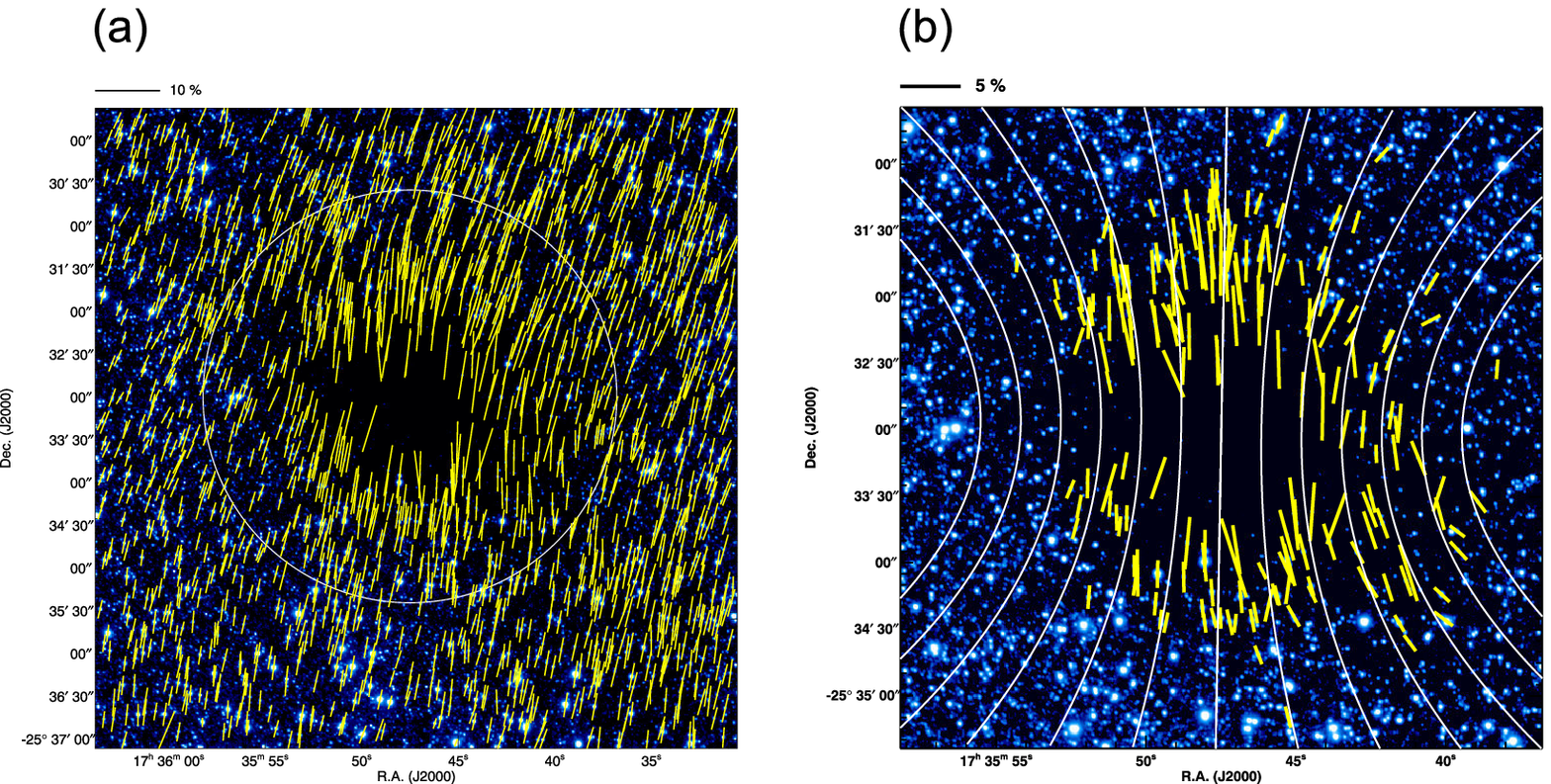}
% \vspace*{-1.0 cm}
 \caption{(a) Polarization vectors of point sources in FeSt 1-457 superimposed on the intensity image in the $H$ band. The white circle shows the core radius (144$''$). The scale of the 10$\%$ polarization degree is shown above the image. (b) Polarization vectors of FeSt 1-457 after subtracting the ambient polarization component. The field of view is the same as the diameter of the core ($288''$ or 0.19 pc). The white lines indicate the magnetic field direction inferred from the fitting with a parabolic function. The scale of the 5$\%$ polarization degree is shown above the image. These figures are taken from Paper I.}
   \label{fig1}
\end{center}
\end{figure}

\clearpage 

\begin{figure}[t]  % \vspace*{-2.0 cm}
\begin{center}
 \includegraphics[width=6.5 in]{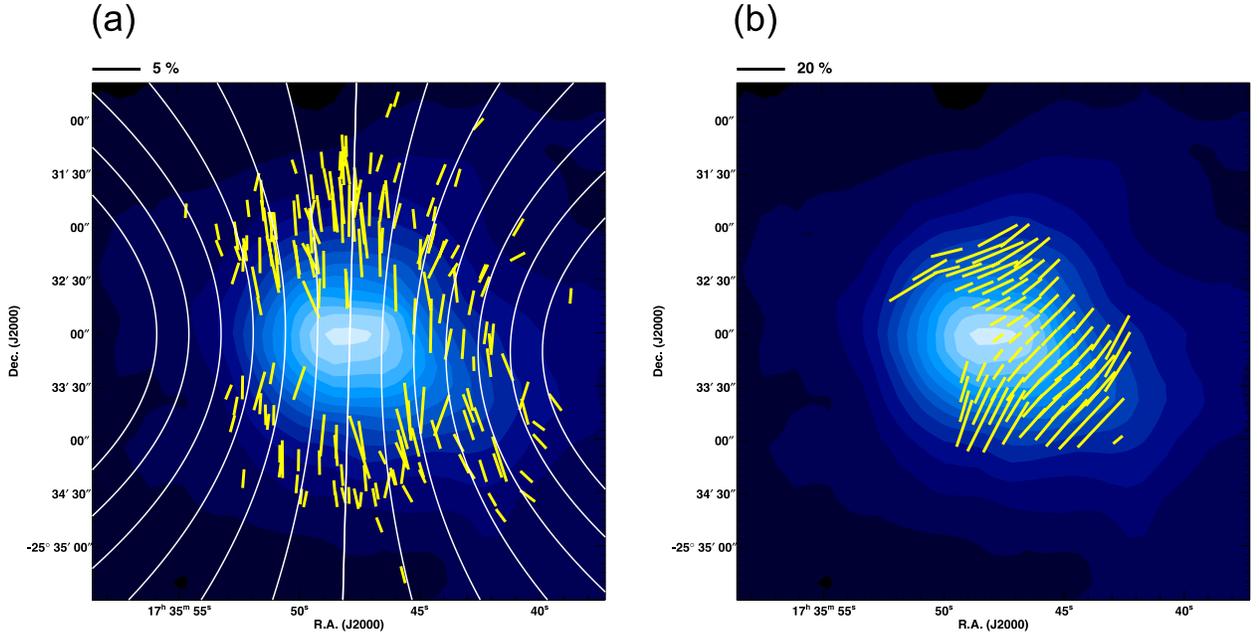}
% \vspace*{-1.0 cm}
 \caption{(a) Polarization vectors of FeSt 1-457 in the $H$ band after subtracting the ambient polarization component, superimposed on the $A_V$ map ($33''$ resolution) taken from Kandori et al. (2005). Filled contour starts from 0 mag with a step of 3 mag. The field of view is the same as the diameter of the core ($288''$ or 0.19 pc). The white lines indicate the magnetic field direction inferred from the fitting with a parabolic function. The scale of the 5$\%$ polarization degree is shown above the image. (b) Polarization vectors of FeSt 1-457 at submm wavelengths (345 GHz) taken from Alves et al. (2014a,b). The polarization vectors were $90^{\circ}$-rotated to show the direction of magnetic fields. The background image is the same as Figure 2a. The scale of the 20$\%$ polarization degree is shown above the image.}
   \label{fig1}
\end{center}
\end{figure}

\clearpage 

\begin{figure}[t]  % \vspace*{-2.0 cm}
\begin{center}
 \includegraphics[width=6.5 in]{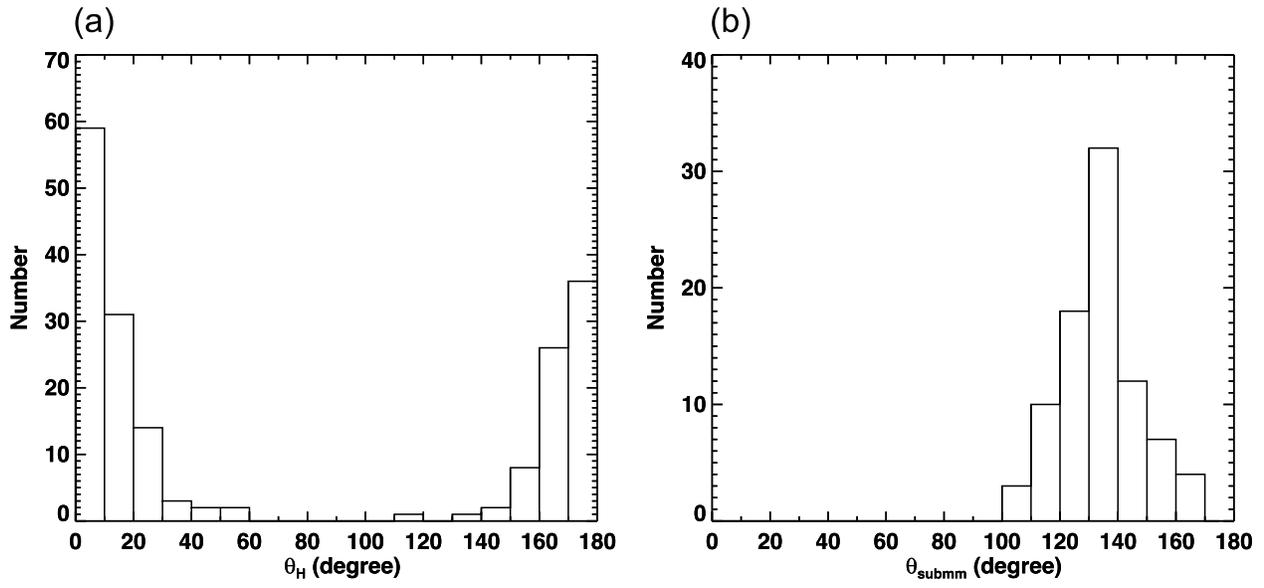}
% \vspace*{-1.0 cm}
 \caption{(a) Histogram of $\theta_H$ for FeSt 1-457 after subtraction of ambient polarization component. (b) Histogram of $\theta_{\rm submm}$ ($90^{\circ}$-rotated) for FeSt 1-457}
   \label{fig1}
\end{center}
\end{figure}

\clearpage 

\begin{figure}[t]  % \vspace*{-2.0 cm}
\begin{center}
 \includegraphics[width=3.0 in]{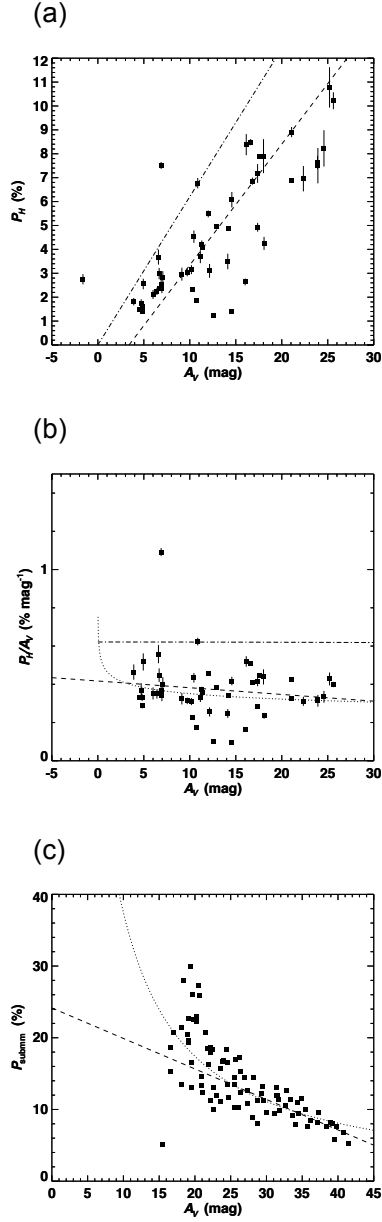}
% \vspace*{-1.0 cm}
 \caption{(a) Relationship between polarization degree at $H$ and $A_V$ toward background stars of FeSt 1-457. Stars with $R \leq 144''$ and $P_H / \delta P_H \geq 10$ are plotted. The relationship was corrected for ambient polarization components, the depolarization effect, and the magnetic inclination angle. The dashed line denotes the linear least-squares fit to the whole data points. The dotted-dashed line shows the observational upper limit reported by Jones (1989). (b) The relationship of the Figure 4a divided by $A_V$ (polarization efficiency). The dashed line denotes the linear least-squares fit to the data points. The dotted line shows the power law fitting result. The dotted-dashed line shows the observational upper limit reported by Jones (1989). (c) Relationship between polarization degree at submm wavelengths and $A_V$. The dashed line denotes the linear least-squares fit to the data points with $A_V \ge 23$ mag. The dotted line shows the power law fitting result.}
   \label{fig1}
\end{center}
\end{figure}

\end{document}